\documentstyle[aps,epsfig]{revtex}
\newcommand{\lb}{l_{\rm B}}
\newcommand{\kt}{k_{\rm B}T}
\begin{document}
\title{Slow plasmon modes in polymeric salt solutions}
\author{I.F. Hakem $^{1,4}$, A. Johner$^{2,3}$ and  T.A. Vilgis$^{1,3}$}
\address{$^{1}$ Max-Planck Institut f\"ur Polymerforschung,
 Ackermannweg 10, D-55122  Mainz, Germany}
\address{$^{2}$ Laboratoire Europ\'een Associ\'e}
\address{$^{3}$ Institut Charles Sadron. 6 rue, Boussingault, F-67083 Strasbourg, France}
\address{$^{4}$ University Aboubakr Belkaid of Tlemcen, Institut of
  Physics, BP 119, Tlemcen 13 000, Algeria}
\date{\today}
\maketitle

\begin{abstract}
  The dynamics of polymeric salt solutions are presented.  The salt consists
  of chains $\rm A$ and $\rm B$, which are chemically different and interact
  with a Flory-interaction parameter $\chi$, the $\rm A$ chain ends carry a
  positive charge whereas the $\rm B$ chain ends are modified by negative
  charges.  The static structure factor shows a peak corresponding to a micro
  phase separation.  At low momentum transfer, the interdiffusion mode is
  driven by electrostatics and is of the plasmon-type, but with an unusually
  low frequency, easily accessible by experiments.  This is due to the polymer
  connectivity that introduces high friction and amplifies the charge
  scattering thus allowing for low charge densities.  The interdiffusion mode
  shows a minimum (critical slowing down) at finite $k$ when the interaction
  parameter increases we find then a low $k$ frequency quasi-plateau.
\end{abstract}

\vspace{1cm}

Scattering methods are a powerful tool to study the structure and phase
behavior of multicomponent polymer mixtures and were the subject of intensive
investigations \cite{BenoitBook}. In general, polymer blends are characterized
by a low entropy of mixing and relatively high enthalpic interactions and
polymers are thus to a large extend immiscible
\cite{Ohta:86.1,Vilgis:90.2,Bates:91.1,Fredrickson:93.1}.  An effective way
to enhance the compatibility of polymer blends, consists to
introduce long range Coulomb forces between particles by adding electrostatic
charges \cite{Brereton:90.1,Vilgis:94.7}.

Quasi elastic scattering experiments allow, in principle, to measure the decay
of various concentration fluctuations in the solution \cite{BenoitBook}.  The
fastest relaxation mode is usually the plasmon mode related to the relaxation
of charge fluctuations.  In the macroscopic limit ($k \to 0$) for small ions,
the plasmon frequency is proportional to the inverse friction coefficient
$\zeta$, and to the Debye-H\"uckel screening parameter $\kappa^{2}$, i.e.,
$\Gamma \propto \kt \kappa^{2} \zeta^{-1}$. Although it is possible to detect
the plasmon by spin echo techniques it remains very difficult to measure,
because of low contrast. Then other processes such as water binding -
unbinding perturbate a clear detection of the plasmon.  To find experimentally
accessible plasmon modes in reasonable time scales larger friction
coefficients, low charge densities, and higher contrast are required.  We
suggest to use polymeric salts, where anions and cations are linked to the end
of two chemically different polymeric backbones.  To be more precise, chains A
and B interact by a Flory - $\chi$ parameter, which allows for
composition fluctuations \cite{DeGennes:79.1}. Many hydrophilic polymers
involve hydrogen bonding and may exhibit a more complex behavior. These
effects can be limited by an appropriate choice of temperature and polymers,
e.g., end-modified POE at room temperature. The chain friction is a power of
$N$ \cite{Doi:86.1} larger than the monomer friction of the order of $\zeta$
introduced above.
\begin{figure} 
       \begin{center} 
       \begin{minipage}{13cm} 
         \centerline{\epsfig{file=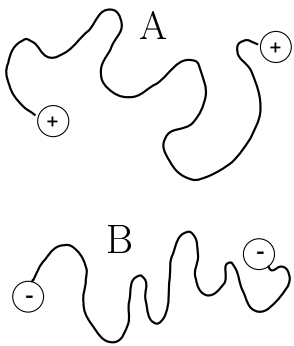,height=4.5cm}}
       \caption {\footnotesize Polycation A and polyanion B in solution.} 
       \end{minipage} 
       \end{center} 
\end{figure}
To benefit from the polymeric effects small ions should be absent, which is
possible in a one to one A-B mixture, this point will be discussed later.
Electroneutrality forbids macrophase separation as $\chi$ increases and
microphase separation at a finite wave vector $k^\star$ is anticipated. Large
scale composition fluctuations violate electroneutrality and their relaxation
is primarily driven by electrostatics. We will indeed show that at low wave
vector the relaxation of composition fluctuations is of the plasmon type but
at an unusually slow time scale. Though critical slowing down closer to the
microphase separation is interesting, we will mainly concentrate on systems
far from the transition where excluded volume effects do not interfere with
the plasmon effects. For the weakly charged chains we have in mind that the
value for $ \kappa$ is small. On the other hand the polymer concentration is
high and high contrast in scattering experiments can be achieved. Symmetric
polymer salt solutions with moderately incompatible backbones (e.g.
hydrogenated and deuterated) are thus good candidates to detect slow plasmon
modes.
       
The static properties of the polymeric salts in equilibrium are well described
by the random phase approximation, since we assume a sufficiently large
overall concentration. The chains are assumed to be of Gaussian connectivity,
the monomers interact with an excluded volume interaction of a strength given
by $V_{ij}$, and the electrostatic interaction between the ionic groups. In
the latter the strength of the Coulomb potential is expressed in term of the
Bjerrum length: $l_{\rm B} = \beta e^2 / 4 \pi \epsilon$, where $e$ and
$\epsilon$ are respectively the electric unit charge and the dielectric
constant. The different chemical nature of the chains is commonly expressed by
a Flory $\chi$ - parameter, $V_{\rm AB} = V + \chi$. For the purpose here it
turns out convenient to use linear combinations of the collective variables to
study the usual plasmon (total charges) modes and the interdiffusion (relative
$\rm A$ and $\rm B$ fluctuations). It is straightforward to calculate the
corresponding static structure factor matrix ${\bf S}({\bf k})$ which is
expressed as a function of the bare structure matrix ${\bf S}_{0} ({\bf k})$,
corresponding to the single chain conformations in the mixture, and the
interaction matrix ${\bf U}_{\rm int} ({\bf k})$ in the form
\begin{equation} 
       \label{Smoins1} 
       {\bf S}({\bf k})^{-1}={\bf S}_{0}({\bf k})^{-1} + {\bf U}_{\rm int}({\bf k}). 
\end{equation} 
The interaction matrix ${\bf U}_{\rm int} ({\bf k})$ contains short (excluded
volume) and long (electrostatic) range interactions.  The elements of the
first matrix are the usual parameters $V_{ij}$.  The relevant structure
factors are ${S^{\rm zz}(\bf k)}$ (charge / charge structure factor), ${S^{\rm
    II}(\bf k)}$ (ionic strength / ionic strength structure factor), ${S^{\phi
    \phi}(\bf k)}$ (concentration / concentration structure factor), ${S^{\rm
    xx}(\bf k)}$ (composition / composition structure factor), ${S^{\rm
    I\phi}(\bf k)}$ (ionic strength / concentration structure factor),
${S^{\rm zx}(\bf k)}$ (charge / composition structure factor).  In the case of
the symmetric salt where chains $\rm A$ are charged positively and $\rm B$
charged negatively, $S^{\rm xx} ({\bf k})$, $S^{\rm zz} ({\bf k})$ and $S^{\rm
  zx} ({\bf k})$ have a peak which increases as the value of the Flory
parameter $\chi$ increases. In this case we find a microphase instability with
a characteristic finite wavevector $k^{*}$.  The microphase separation at
non-vanishing vector $k^*$ occurs for a sufficiently strong incompatibility.
For $\chi = 0$ or $\chi$ finite but sufficiently small, entropy effects are
dominant and favor mixing.

In light or neutron scattering experiments, one can measure directly the
matrix of dynamical functions $S_{ij} ({\bf k}, t)$. To compute the relevant
elements of the dynamic structure factor matrix we start from the generalized
Langevin equation, which describes the (short) time evolution of ${\bf S}({\bf
  k}, t)$ and, neglecting memory effects, lead us to a single  exponential
decay of ${\bf S}({\bf k}, t)$, ${\bf S}({\bf k}, t)={\bf e}^{-{\bf \Omega}
  ({\bf k}) t} {\bf S}({\bf k})$.  The decay rate, which is experimentally
available is given by the first cumulant matrix $\Omega ({\bf k})$ which is
related to different correlation functions. Hydrodynamic interactions between
monomers are taken into account at the Oseen level
\cite{Doi:86.1,Vilgis:91.1}.
\begin{equation} 
\label{omega}
       {\bf \Omega} ({\bf k}) = k_{\rm B} T 
       \int{{\mbox{d}{\bf q}\over(2\pi)^3}{\bf S}( 
       {\bf k + q}){k^2 - ({\bf q \cdot k}/q)^2\over \eta q^2}{\bf S}^{-1}({\bf k})} 
       \label{kubo} 
\end{equation} 
The eigenvalues of ${\Omega (\bf k)}$ lead us to the decay rates of thermal
fluctuations in the solution. In a first step the eigenvalues of the frequency
matrix eq.(\ref{omega}) have been computed numerically.

Let us first concentrate on the fast (usual) plasmon mode, which corresponds
to the relaxation of charge fluctuations. At high wavevector limit, the
relaxation is diffusive as for a monomeric salt and the electrostatic
interactions are irrelevant. In the low $k$ limit, the symmetric salt
behaves as an usual monomeric salt. The relaxation of charge fluctuations of
size larger than $\kappa^{-1}$ is driven by electrostatics.  Since the chain
ends are different from the rest of the chain monomers, an additional
relaxation mode is expected.  At high wave vectors, the relaxation is
diffusive and at larger length scales, the relaxation is rather driven by
chain elasticity.  The slowest mode corresponds to the relaxation of monomer
concentration fluctuations and is similar to the mode observed in binary
polymer-solvent solutions. This mode is independent of the strength of the
electrostatic interaction and, depends mainly on the excluded volume $V$. It
is insensitive to the Flory parameter $\chi$. At small length scales, the
relevant friction is proportional to the probed length scale $1/k$ and in the
limit of the large length scales, the relevant friction is proportional to the
correlation length $\xi$.

Let us now turn to the interdiffusion mode. If the charges were absent the
interdiffusion would show the classical critical slowing down of the
composition fluctuations close to the macro phase separation. Charges
stabilize the solution against macrophase separation and the relaxation of
composition fluctuations at low $k$ is dominated by electrostatics: the
relaxation frequency has a finite value in the macroscopic limit. This
behavior is characteristic for plasmon modes. As can be seen on the Fig. 2
below the frequency is very low (of the order of $10^4$Hz, for the typical
values shown). At wavevector zero a weak (not visible at the scales of Fig. 2)
splitting occurs, due to the dependence of the mobility on the Flory-$\chi$
parameter.  The behavior found here is somewhat similar to block copolymers.
The usually observed critical slowing down is shifted to the local scale
$k^\star$. Thus the corresponding relaxation frequency shows a minimum for
high enough values (higher than chosen in the figure) of the interaction
parameter $\chi$.
\begin{figure} 
 \begin{center} 
   \begin{minipage}{13cm} 
       \centerline{\epsfig{file=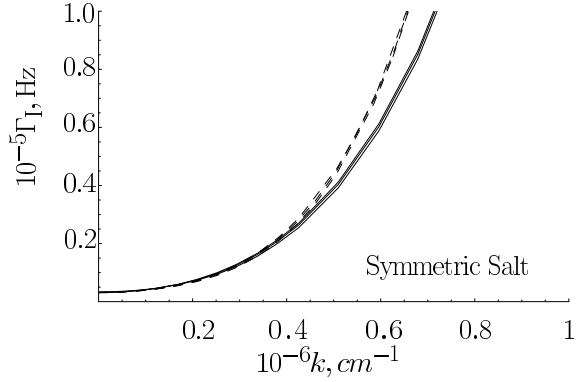,height=5cm}} 
       \caption {\footnotesize Variation of Interdiffusive mode versus the 
         wave vector ${\bf k}$. Dashed curves correspond to system (2X2)
         Continuous curves correspond to system (4$\times$4) for different values of
         $\chi$ ( 0, 0.1 and 0.24 ), $N=10 ^ 4, b=\lb= 7 10^{-8}cm, \phi a^3 /
         N = 3.67 10^{-6}.$}
  \end{minipage} 
 \end{center} 
\end{figure} 
The analytical calculation of the full $4 \times 4$ system involves tedious algebra and
produces long and non transparent formulae. In the low wavevector limit the
precise position of the charges is not of specific interest \footnote{This
  statement is less obvious for the mobility integral occurring in
  eq.(\ref{kubo}). However it can be checked afterwards that the integral is
  not dominated by the large $k$ range.}.  Therefore it is convenient to smear
out the charges.  Thus it is possible to replace the system described above by
an effective two chain model with a charge $f=2/N$ per monomer.  The static
composition structure factor $S_{\rm xx}$ can then be represented as
\begin{equation} 
  \label{xx1} 
   S_{\rm xx} = \frac{\phi N}{2} \left\{ 
   \frac{1}{\left(2 - \chi \phi N \right) + k^2 R_{\rm g}^2 + 
       4 \kappa^2 /k^2 } 
       \right\} , 
\end{equation} 
where $\kappa^2 = 4\pi \lb (2\phi)/N$. If the chains are uncharged, the
classical case of a binary polymer blend is recovered. The term proportional
to $1/k^{2}$ stems from the bare Coulomb interaction between the charges.
$S_{\rm xx}$ diverges then at ${\bf k=0}$ and $\chi \phi N = 2$.  The charges
themselves enhance compatibility and yield a microphase separation transition
at $\chi \phi N = 2 + 4 \kappa R_{\rm g}$. The structure factor $S_{\rm xx}$
shows a characteristic peak at a finite wavevector $k^*$ where $k^{\star 2} =
2\kappa/R_g$.  For understanding the dynamical results obtained by numerical
computation of the full four component system it is further convenient to
rewrite the structure factor to
\begin{equation} 
       \label{xxq} 
       S_{\rm xx} = {\phi N \over 2} {1\over \sqrt{\Delta}} 
       \left \{ 
       \frac{1}{1+k^2 \xi_1^2} - \frac{1}{1+k^2 \xi_2^2} 
       \right \} 
\end{equation} 
where we have defined the discriminant $\Delta = (2-\chi \phi N)^2 - (4 \kappa
R_{\rm g})^2$.  In this representation two important length scales show up,
i.e., $\xi_{1,2}^2 = (2R_{\rm g}^2)/(2-\chi \phi N \pm \sqrt{\Delta})$. It is
immediately seen that $\xi_1 \xi_2 = k^{*-2}$. For weak incompatibility, i.e.,
$\chi \phi N \leq 2 - 4 \kappa R_{\rm g}$, the correlation function in real
space consists of two exponential decaying functions $\exp(-r/\xi_{1,2})$, see
eq (\ref{xxq}). Closer to the microphase instability, $2 - 4\kappa R_{\rm g} <
\chi \phi N < 2 + 4\kappa R_{\rm g}$, the real space structure factor contains
decaying oscillatory parts. In the latter regime the plasmon behavior of the
composition mode at low $\bf k$ will turn out to be hidden by the composition
fluctuations.  Moreover renormalization theories beyond the RPA are required.
   
The dynamics can be understood within the same framework. The relaxation
frequencies can be computed along the lines discussed above. It is sufficient
to understand the behavior of the relaxation frequencies from figure (1) at
low values for the wavevector.  Most interesting is the limit at zero
wavevectors
\begin{equation} 
       \Omega_{\rm xx}({\bf k = 0}) = 
       \frac{\kt}{6\pi \eta} k^{\star 4} \left( {1\over \xi_{1}} + {1\over 
       \xi_{2}}\right)^{-1}, 
\end{equation} 
which has to be discussed in several limits. The first is the limit of
$\chi=0$, i.e., no composition fluctuations and therefore no coupling between
thermodynamic and electrostatic interactions. For this case we find the simple
result
\begin{equation}
       \label{chinull} 
       \Omega_{\rm xx}(\chi=0) = {\kt \over 6\pi \eta} \;\; \frac{4 
       \kappa^{2}}{R_{\rm g}\sqrt 2}. 
\end{equation} 
where $\kappa R_g << 1$ was taken into account. The physical interpretation
here is indeed simple: The relaxation frequency corresponds to a plasmon mode
of $\phi / N$ macroions carrying two elementary charges each, whose
hydrodynamic interaction is of the range of the radius of gyration $R_{\rm g}$
(corresponding to the Oseen - tensor $1/r$). Composition fluctuations do not
couple in this limit, thus the non appearance of any wavevector $k^{*}$ is
natural, in contrast to the upper boundary $\chi \phi N = 2 - 4\kappa R_{\rm
  g}$. There the first sign of the micro structure shows up in terms of
$k^{*}$. Indeed we find
\begin{equation} 
\label{limits}
       \Omega_{\rm xx}(\chi\phi N = 2 - 4\kappa R_{\rm g}) = {\kt \over 6\pi \eta}\left( k^{*}\right)^{3}/2 
\end{equation} 
In this regime of the interaction parameter the dynamics is ruled by the
intrinsic length scale $k^\star$ of the microphase separation transition.  The
plasmon character of the composition mode is reflected by eq. (\ref{chinull}).
For its observation a weak incompatibility like between deuterated and hydrogenated
polymers is necessary.  In the other limit which is given by eq.(\ref{limits})
the plasmon character is already blurred by the microphase separation.  The
slow plasmon mode is predicted in a range accessible by quasi elastic light
scattering and spin echo techniques.  The dynamical study has been limited to
short time scales where the decays are exponential. Entanglements have been
ignored completely. With respect to our studies, they will only further slow
down all polymeric modes.  The measurements are carried out in aqueous
solution, and small amount of salt and ions could destroy this behavior. We
carried out a model calculation, which involves these ions as individual
component in the Langevin dynamics.  The small ions couple to the electric
charges bound at the chains, via a Coulomb potential. The dynamics of these
ions with respect to all polymer modes involving the composition and
concentration is a power of $N$ faster, $N$ being the chain length. Therefore
the fast small ions produce  a screened electrostatic potential of the Debye -
H\"uckel type, where the screening constant is determined by the concentration
of the remaining small ions. The relevant concentration in water is often of
order 10$^{-6}$, slightly higher than imposed by self-dissociation, and the
corresponding wave vector is about $k = 10^{4}$ cm$^{-1}$. Thus we expect that
at larger wave vectors the slow plasmon mode can be observed.

Acknowledgement: The funding of this work by the Deutsche
Forschungsgemeinschaft, Max-Planck-Society and L.E.A. are most gratefully
acknowledged. We thank Henri Benoit, Michel Rawiso, Jean-Fran{\c c}ois
Joanny, Mustapha Benmouna and Burkhard D\"unweg for many helpful discussions.

\end{document}